\documentclass[12pt]{article}
\usepackage{epsfig,amssymb,euscript,bbold}
\usepackage{amsmath}
\def\be#1\ee{\begin{equation}#1\end{equation}}
\newcommand{\bea}{\begin{eqnarray}}
\newcommand{\eea}{\end{eqnarray}}
\newcommand{\ba}{\begin{array}}
\newcommand{\p}[1]{(\ref{#1})}
\newcommand{\ea}{\end{array}}

\def\bbox{{\,\lower0.9pt\vbox{\hrule \hbox{\vrule height 0.2 cm
\hskip 0.2 cm \vrule height 0.2 cm}\hrule}\,}}
\newcommand{\dsl}{\pa \kern-0.5em /}

\newcommand{\nn}{\nonumber \\}

\newcommand{\diag}{{\rm diag}\,}

\newcommand{\dcy}{d}               % dimension of cycle
\def\ds{\raise.15ex\hbox{/}\kern-.57em\partial}
\def\Ds{\,\raise.15ex\hbox{/}\mkern-13.5mu D}
\font\mybb=msbm10 at 10pt
\def\bb#1{\hbox{\mybb#1}}

\def\bR {\bb{R}}

\def\bC {\bb{C}}

\begin{document}

\makeatletter
\renewcommand{\theequation}{\thesection.\arabic{equation}}
\@addtoreset{equation}{section}
\makeatother

\baselineskip 18pt

%%%%%%%%%%%%%%%% title page %%%%%%%%%%%%%%%%%%%%%%%%%%%%%%%%%%%%

\begin{titlepage}
\vfill
\begin{flushright}
QMUL-PH-02-04\\
hep-th/0202184\\
\end{flushright}

\vfill

\begin{center}
\baselineskip=16pt
{\Large\bf M-Theory solutions with AdS factors\\}
%\Large\bf without supersymmetry}
\vskip 10.mm
{Jerome P. Gauntlett$^{*\ddag 1}$, ~Nakwoo Kim$^{\dag 2}$, 
Stathis Pakis$^{*3}$ 
and Daniel Waldram$^{*\ddag 4}$}\\
\vskip 1cm
%\vfill
$^*$ {\small\it 
Department of Physics\\
Queen Mary, University of London\\
Mile End Rd, London E1 4NS, U.K.\\}
\vskip 0.5cm
$^\dagger${\small\it Max-Planck-Institut f\"ur Gravitationsphysik\\
Albert-Einstein-Institut\\
Am M\"uhlenberg 1, D--14476 Golm, Germany\\}
\vskip 0.5cm
$^\ddag${\small\it Isaac Newton Institute for Mathematical Sciences\\
University of Cambridge\\
20 Clarkson Road, Cambridge, CB3 0EH, U.K.\\}
\vspace{6pt}
\end{center}
\vfill
\par
\begin{center}
{\bf ABSTRACT}
\end{center}
\begin{quote}
Solutions of $D=7$ maximal gauged supergravity are constructed
with metrics that are a product of a $n$-dimensional anti-de Sitter ($AdS$) 
space, with $n=2,3,4,5$, and certain Einstein manifolds. 
The gauge fields have the same form as in the recently constructed 
solutions describing the near-horizon limits of M5-branes 
wrapping supersymmetric cycles. 
The new solutions do not preserve any 
supersymmetry and can be uplifted to obtain new 
solutions of $D=11$ supergravity, which are warped and
twisted products of the $D=7$ metric with a squashed four-sphere. 
Some aspects of the stability of the solutions are discussed.
\vfill
\vskip 5mm
\hrule width 5.cm
\vskip 5mm
{\small
\noindent $^1$ E-mail: j.p.gauntlett@qmul.ac.uk \\
\noindent $^2$ E-mail: kim@aei-potsdam.mpg.de\\
\noindent $^3$ E-mail: s.pakis@qmul.ac.uk \\
\noindent $^4$ E-mail: d.j.waldram@qmul.ac.uk \\
}
\end{quote}
\end{titlepage}
%%%%%%%%%%%%%%%%%%%%%%%%%%%%%%%%%%%%%%%%
\setcounter{equation}{0}

\section{Introduction}

A rich class of supersymmetric solutions of $D=10$ and $D=11$ supergravity
have recently been found that describe the near-horizon limits of branes
wrapping supersymmetric cycles \cite{malnun}--\cite{apreda}. 
These solutions are dual to the supersymmetric field theories arising on the
branes. There are two key features of the construction of the
supergravity solutions. First, they are initially constructed in
a suitable gauged supergravity and then subsequently uplifted to
$D=10$ or $D=11$. Secondly, the ansatz for the gauge fields
in the gauged supergravity are determined by the geometry of
the normal bundle of the supersymmetric cycle. This latter feature
is a manifestation of the fact that the field theories arising on
the wrapped branes are coupled to external R-symmetry currents,
or ``twisted'', in order to preserve supersymmetry \cite{bvs} .

For almost all cases of M-fivebranes wrapping $d$-dimensional
supersymmetric cycles, $D=11$ supergravity solutions were found that
include an $(7-d)$-dimensional anti-de~Sitter factor
\cite{malnun,agk,gkw,gk}. More precisely, the solutions are warped and
twisted products of an anti-de~Sitter factor, a cycle with an Einstein
metric and a squashed four-sphere. The $AdS$ factor indicates that, in
the infrared, at length scales much larger than the size of the cycle, the
corresponding $(6-d)$-dimensional field theory on the wrapped
M-fivebranes is super-conformal, a fact which {\it a priori} was not
at all clear.  

Here we would like to report on a new class of solutions of
$D=11$ supergravity with $AdS$ factors that preserve no
supersymmetry. They are obtained in maximal $D=7$ gauged
supergravity using exactly the same ansatz that
was used to find the supersymmetric wrapped M5-brane solutions, and 
consequently they have a similar $D=11$ structure. In the supersymmetric
$AdS$ solutions the Einstein metric on the cycle typically has negative
curvature. In the new non-supersymmetric solutions we find a more
democratic mixture of both negatively and positively curved cycles.

Our main interest in these solutions is that they
might provide M-theory duals of new conformal field
theories with no supersymmetry. Since the $AdS$ factors that appear
range from $n=5$ to $n=2$, these would be dual to conformal field
theories in dimensions ranging from four to one. The most
interesting case could be the dualities between the
$AdS_5$ solutions and four-dimensional field 
theories. In this context, to our knowledge, the only previously known
solutions of this type are the compactifications on
K\"{a}hler--Einstein six-dimensional manifolds 
given in~\cite{pvann}. A necessary condition for the correspondence is
that the solutions be stable. We have checked that certain
perturbations of the scalar fields satisfy the Breitenlohner-Freedman
bound \cite{bf1,bf2,mt} in most cases, but we will 
leave a full investigation of this issue to future work. We note that
we do not know of any stability analysis that has been performed
for the solutions in~\cite{pvann} just mentioned.
Additional instabilities might arise because of the presence of 
massless fields \cite{berkone,berktwo}.
Furthermore, one also needs to establish that there are non-perturbative 
instabilities that could be of the general form discussed in \cite{dggh,mms}. 
We shall not attempt to address
these issues here. Assuming that the correspondence is valid, we note that
the central charges of the conformal field theories would be proportional
to $N^3$ as in the supersymmetric cases \cite{malnun,agk,gkw,gk}.

The plan of the paper is as follows. In section~2 we describe the essential
aspects of the ansatz that we employ to find the new solutions in $D=7$ gauged
supergravity. As noted, the ansatz includes a $d$-dimensional cycle with an
Einstein metric and this is used to consistently truncate the theory,
via Kaluza-Klein reduction on the cycle, to an effective $(7-d)$-dimensional
theory of gravity coupled to two scalar fields. (The exception is the
case of five-cycles, i.e. $d=5$, where we find the solutions directly 
from the $D=7$ equations of motion.)
Some general comments concerning the stationary 
points and their stability in this reduced
theory then precedes a more detailed description of the different cases in 
the following sections. In section~3, 4, 5 and~6 we discuss the
$AdS_5$, $AdS_4$, $AdS_3$ and $AdS_2$ solutions, respectively. 
The different cases are labelled by the kind of supersymmetric cycle 
that a fivebrane can wrap and for completeness we have 
included the supersymmetric $AdS$ solutions already 
found in \cite{malnun,agk,gkw,gk}.
The paper concludes in section~7, where we have 
included a table summarising both the supersymmetric and non-supersymmetric
solutions. We also comment on the possible connection between our new 
non-supersymmetric solutions and the supersymmetric conformal 
field theories arising on M5-branes wrapping supersymmetric cycles.

%%%%%%%%%%%%%%%%%%%%%%%%%%%%%%%%%%%%%%%%%%%%%%%%%%%%%%%%%%%%%%%%%%%%%%%%%%%

\section{Effective actions and obtaining AdS solutions}
\label{7dsugra}

\subsection{Maximal $D=7$ gauged supergravity and ansatz}
\label{sec:ansatz}

The Lagrangian for the bosonic fields of maximal gauged supergravity
in $D=7$ is given by \cite{vn}
\begin{equation}
\begin{aligned}
  {\cal L} &= \sqrt{-g} \bigg[ R 
         + \frac{1}{2}m^2\left(T^2-2T_{ij}T^{ij}\right)
         - P_{\mu ij}P^{\mu ij} 
         - \frac{1}{2} \left({\Pi_A}^i{\Pi_B}^jF_{\mu\nu}^{AB}\right)^2
      \\ & \qquad
         - m^2 \left({{\Pi^{-1}}_i}^AS_{\mu\nu\rho,A} \right)^2
      \bigg]
      - 6m \delta^{AB}S_A\wedge F_B 
      \\ & \qquad
      + \sqrt{3}\epsilon_{ABCDE}
          \delta^{AG}S_G\wedge F^{BC}\wedge F^{DE}
      + \frac{1}{8m} \left(2\Omega_5[B]-\Omega_3[B]\right)
\end{aligned}
\label{Ssugra}   
\end{equation}
Here $A,B=1,\dots,5$ denote indices of the $SO(5)_g$ gauge-group, while
$i,j=1,\dots,5$ denote indices of the $SO(5)_c$ local composite
gauge-group, which are raised and lowered with $\delta^{ij}$ and
$\delta_{ij}$. The 14 scalar fields ${\Pi_A}^i$ are given by the coset
$SL(5,\bR)/SO(5)_c$ and transform as a ${\bf 5}$ under both $SO(5)_g$
(from the left) and $SO(5)_c$ (from the right). The term that gives
the scalar kinetic term, $P_{\mu ij}$, and the $SO(5)_c$ composite
gauge field, $Q_{\mu ij}$, are defined as the symmetric and
antisymmetric parts of ${(\Pi^{-1})_i}^A\left({\delta_A}^B \partial_\mu + 
2m B_{\mu\,A}{}^B\right){\Pi_B}^k \delta_{kj}$, respectively. 
Here $B_A{}^B$ are the $SO(5)_g$ gauge fields with field strength 
$F^{AB}=\delta^{AC}{F_C}^B$, and note that the gauge coupling constant
is given by $2m$. The four-form field strength $F_A$ for the three-form
potential $S_A$, is given by 
the covariant derivative $F_A=dS_A+2mB_A{}^BS_B$. The
potential terms for the scalar fields are expressed in terms of
$T_{ij}={{\Pi^{-1}}_i}^A{{\Pi^{-1}}_j}^B\delta_{AB}$ with
$T=\delta^{ij}T_{ij}$. Finally, $\Omega_3[B]$ and  $\Omega_5[B]$ are
Chern-Simons forms for the gauge fields $B$ that will not play a role
in this paper.  We use a ``mostly plus'' signature convention for the
metric.
The supersymmetry transformations of the fermions
in these conventions were given in \cite{gkw}.

The new solutions we find here are obtained from the same ansatz used
in~\cite{gkw} and~\cite{gk}. The geometry is taken to be a product of
a $(7-d)$-dimensional space with a $d$-dimensional cycle $\Sigma_d$, so
that   
\begin{equation}
\label{metansatz}
   ds^2 = e^{-2d\phi} ds^2_{7-d} 
      + e^{2(5-d)\phi} d\bar{s}^2(\Sigma_d)
\end{equation}
where $ds_{7-d}^2$ is the metric on the uncompact space, and
$d\bar{s}_d^2$ is the metric on the $\dcy$-cycle, $\Sigma_d$. The
scalar field $\phi$ is assumed to depend on the coordinates on
$ds_{7-d}^2$ and the exponents are chosen so that the
reduced $(7-d)$-dimensional effective action has a conventional Einstein
term.  (Note that in section~\ref{sec:slag4prod} where we discuss
the geometry arising from fivebranes wrapped on a product of cycles
this metric ansatz is generalised slightly, to allow for a different
conformal factor for each cycle. However, it is still precisely the
same ansatz used in~\cite{gk}.) Ultimately, we will be interested in
solutions where $\phi$ is a constant and the $(7-d)$-dimensional space
is $AdS$. For each case we will present the $AdS$ radius for the
line element $ds_{7-d}^2$ without the conformal factor, as this is
proportional to the central charge of the putative dual conformal
field theory. Our definition of the radius $R$ of $AdS_n$ is given by 
\be\label{Rdef}
R_{\mu\nu}=-\frac{n-1}{R^2}g_{\mu\nu}
\ee
Again following the ansatz for supersymmetric solutions, the metric on
the $\Sigma_d$ is assumed to be at least Einstein, satisfying  
\be\label{ein}
   \bar{R}_{ab} = l m^2 \bar{g}_{ab}
\ee
where we can always rescale so that $l=0,\pm 1$. (In fact, we will
only find $AdS$ solutions for the cases $l=\pm 1$.) The Einstein
condition implies that the Riemann tensor can be written 
\be\label{decom}
\bar R_{abcd}=\bar C_{abcd} +\frac{2lm^2}{d -1}\bar g_{a[c}\bar g_{d]b}
\ee
where ${\bar C}$ is the Weyl tensor and is only present for $d\ge 4$. 
The additional conditions that are placed on ${\bar C}$ will be discussed
case by case later.
Note that, for comparison with conventions used in \cite{gkw,gk}, we have
$e^{2(5-d)\phi}=m^2 e^{2g}$.

The ansatz for the $SO(5)$ gauge fields incorporates the twisting required
for supersymmetric solutions and is specified by the spin
connection with respect to the metric on $\Sigma_d$. 
In general, we will decompose the $SO(5)$ symmetry into $SO(p)\times
SO(q)$ with $p+q=5$, and excite the gauge fields in the $SO(p)$
subgroup. The precise form was given in~\cite{gkw,gk} and
will be summarized case by case below.  

In order to respect the $SO(p)\times SO(q)$
decomposition, the ansatz for the scalar fields,  
is restricted to a single scalar mode:
\be\label{scalaransatz}
{\Pi_A}^i=\diag(e^{q\lambda},\dots,e^{q\lambda},
e^{-p\lambda},\dots,e^{-p\lambda})
\ee
where we have $p$ followed by $q$ entries. This implies that the
composite gauge field $Q$ is then determined by the gauge fields via
$Q^{ij}=2m B^{ij}$. 

The form of the three-form potentials $S_A$ and four-form field
strengths $F_A$, is, in general, determined by the gauge fields and
scalar fields, via the $S_A$ equation of motion. As discussed case by
case in~\cite{gkw,gk}, there are two distinct classes of ansatz. For
$d\leq 3$ and two cases with $d=4$, one can always set
$S_A=F_A=0$. For the remaining cases with $d=4$ and for $d=5$, it is
consistent to take $F_A=0$, but now $S_A$ is non-zero. In particular,
for the $d=4$ cases we always have $p=4$ and $q=1$, so that the
$SO(5)$ index $A$ is naturally labelled via the split $A=(m,5)$ where
$m=1,\dots,4$. We then have $S_m=0$ while 
\begin{equation}
   S_5=-\frac{1}{2\sqrt{3}}\,c\,e^{-8\lambda-4\phi}
         e^0 \wedge e^1 \wedge e^2
\label{eq:Sansatz}
\end{equation}
where $(e^0,e^1,e^2)$ are an orthonormal frame for $ds_{7-d}^2$, and
the constant $c$ is given by 
\begin{equation}
   \label{eq:cdef}
   c = \frac{1}{8m^2}\epsilon^{abcd}\epsilon_{mnpq}
          F_{ab}^{mn}F_{cd}^{pq}
\end{equation}
where $a,b=1,\dots,4$ denote coordinates on $\Sigma$. (Note the
normalisation of $c$ is slightly different from that used
in~\cite{gkw,gk}.) 
%\textbf{[$c$ here is Nakwoo's $c/m^4$]}. 
The value
of $c$ depends, through $F^{mn}$, on the ansatz for the gauge fields,
and so changes case by case. The $d=5$ case will be discussed
separately in section~\ref{sec:ads2}.

Note that much of the structure of this ansatz~\cite{gkw,gk},
following~\cite{malnun}, arose from requiring the solutions to be
supersymmetric. Given this is no longer a requirement, there are, of
course, a number of generalisations one might consider. One very
simple possibility when $q>1$, is to break the $SO(q)$ symmetry in the
scalar field space so that we now have a set of scalars
$\lambda_1,\dots,\lambda_q$ with  
\be\label{genscalaransatz}
{\Pi_A}^i=\diag(e^{\sum\lambda_i},\dots,e^{\sum\lambda_i},
e^{-p\lambda_1},\dots,e^{-p\lambda_q})
\ee
Actually, this possibility could {\it a priori} be consistent 
with supersymmetry as it does not destroy the twisting.
We note that we did, in fact, consider generalisations of this
type. We find that this leads to one new $AdS$ solution beyond
those found using the simpler ansatz~(\ref{scalaransatz}). Note that
this analysis revealed that a solution presented in \cite{pernicisezgin} 
is not in fact a solution to the equations of motion.

The solutions of $D=7$ gauged supergravity containing $AdS$ factors
that we obtain via this ansatz can then be used to obtain solutions to
$D=11$ supergravity using the uplifting formulae given
in~\cite{vn,vntwo,vnthree}. In particular the $D=11$ metric takes the
form 
\begin{equation}
\begin{split}
   \label{eq:uplift}
   ds^2_{11} &= \Delta^{-2/5} \left[ e^{-2d\phi} ds^2_{7-d} 
            + e^{2(5-d)\phi} d\bar{s}^2(\Sigma_d) \right] \\
        &\qquad \qquad
            + m^{-2} \Delta^{4/5} \left[
               e^{2q\lambda}DY^a DY^a + e^{-2p\lambda}dY^idY^i \right]
\end{split}
\end{equation}
where 
\begin{equation}
\begin{aligned}
   DY^a &= dY^a + 2m B^{ab}Y^b \\
   \Delta^{-6/5} &= e^{-2q\lambda}Y^aY^a + e^{2p\lambda}Y^iY^i
\end{aligned}
\end{equation}
The indices run over $a=1,\dots,p$ and $i=p+1,\dots,5$, and
$(Y^a,Y^i)$ are constrained coordinates on $S^4$ satisfying
$Y^aY^a+Y^iY^i=1$. The presence of non-zero $\lambda$ and
$B^{ab}$ mean that the sphere is squashed and twisted. The four-form
field strength of the eleven-dimensional supergravity is
proportional to the volume form on the squashed $S^4$ together with
additional terms due to the gauging. Its precise form is given
in~\cite{vntwo,vnthree}.

\subsection{Truncated effective theory}
\label{sec:truncated}

In all but the $d=5$ case which will be dealt with separately in
section~\ref{sec:ads2}, our procedure for finding solutions is as
follows. Using the fact that in our ansatz the $(7-d)$-metric and the
two scalars $\phi$ and $\lambda$ do not depend on the co-ordinates
of the cycle $\Sigma_d$, we first truncate the seven-dimensional gauged
supergravity to obtain an effective $(7-d)$-dimensional theory based 
on these fields. We then look for stationary points of the effective
potential for the scalar fields, which in all cases turn out to be
at points where $V$ is negative. Thus these correspond to solutions
where the $(7-d)$-dimensional space is $AdS$. 
%As we discuss these solutions may or may not be stable. 

To be explicit, after truncation,
the effective Lagrangian in $7-d$ dimensions is given by
%substituting our ansatz in the seven-dimensional
%action~(\ref{Ssugra}), and integrating over the cycle $\Sigma_d$, we
%find an effective Lagrangian in $7-d$ dimensions given by
%
\begin{equation}\label{eq:Seff}
   \mathcal{L} = \sqrt{-g} \Big[ R 
        - 5d(5-d) \left(\partial\phi\right)^2
        - 5pq \left(\partial\lambda\right)^2
        - V(\phi,\lambda) \Big]
\end{equation}
where $g_{\mu\nu}$ is the metric for the $(7-d)$-dimensional
line element $ds_{7-d}^2$ appearing in~(\ref{metansatz}). 
The first two terms arise from the reduction
of the seven-dimensional 
curvature $R$ and scalar kinetic $P^2$ terms in~(\ref{Ssugra}). The
effective potential $V$ comes from the seven-dimensional curvature
$R$ and the remaining terms and is given by  
\bea
\label{eq:Veff} 
   m^{-2}\,V(\phi,\lambda) &=& - \frac{1}{2}e^{-2d\phi} \left[
           p(p-2) e^{-4q\lambda} + 2pq e^{2(p-q)\lambda}
           + q(q-2)e^{4p\lambda} \right]
       \nn && \quad 
           -\; dle^{-10\phi}
           + ke^{4q\lambda+2(d-10)\phi}
           + \frac{1}{2}c^2e^{-8\lambda-16\phi} 
\eea 
where 
\be
\label{eq:kdef}
   k = \frac{1}{2m^2} 
          \bar{g}^{ac}\bar{g}^{bd}F_{ab}^{mn}F_{cd}^{mn}
\ee
and so depends on the ansatz for the gauge fields. The effective
potential depends on the integers $d$, $p$ and $q=5-p$, the sign $l$
of the curvature of the Einstein metric on the cycle, as well as the
constants $c$ and $k$, all of which vary case by case. It is useful to
distinguish between those cases with $c=0$ and those with $c\neq
0$. Calculating $k$ in each case, one can show that, in general,  
\begin{equation}
\begin{aligned}
   \text{if $d=4$ and $p=4$:} &&\quad 
      c &\neq 0,  &\quad  k &= c \\
   \text{otherwise:} &&\quad 
      c &= 0,  &\quad  k &= \frac{d(2d+2p-dp)}{8p}
\end{aligned}
\label{eq:kc}
\end{equation}
Note that, as usual, one must derive this action via substitution of
the ansatz into the equations of motion. Substitution directly into
the action can lead to the wrong equations of motion. In particular,
here one would obtain the wrong sign for the final term
in~(\ref{eq:Veff}). More generally we have in fact shown that
this reduction is in fact consistent. In other words,
that {\it any} solution to the equations of motion of the reduced
theory give rise to a solution of $D=7$ gauged supergravity.

Before turning to the general problem of finding the stationary points
of $V$, note that, since we know that the ansatz admits supersymmetric
solutions, we might expect that the effective action has a simple
supersymmetric generalisation. 
In particular, it may well be possible to extend the ansatz we are
considering to obtain a consistently truncated supersymmetric theory. 
We shall leave such an investigation to future work, but let us note that we
can recast the effective potential in terms of a putative
superpotential $W$ as follows:
\begin{equation}
   \frac{1}{2} V = K^{ab} \left( \partial_a W \partial_a W\right) - \beta^2 W^2
   \label{eq:VfromW}
\end{equation}
where $a$ and $b$ label scalar fields $A^a=(\lambda,\phi)$, $K^{ab}$
is the inverse of the sigma-model metric for the scalar kinetic terms
$2K_{ab}\partial A^a\partial A^b= 5d(5-d)(\partial\phi)^2 +
5pq(\partial\lambda)^2$ and the constant $\beta^2$ depends on the
dimension of the effective theory and is given by
\begin{equation}
\label{eq:beta}
   \beta^2 = \frac{2(6-d)}{5-d}
\end{equation}
Explicitly, we have
\be 
   W = \frac{1}{4}me^{-d\phi}\left(
              pe^{-2q\lambda}+qe^{2p\lambda}\right)
       + \frac{1}{8}mdle^{2q\lambda+(d-10)\phi}
       - \frac{1}{4}mc e^{-4\lambda-8\phi} 
\ee 
provided that $k$ and $c$ are given as in~\eqref{eq:kc}. 
For the four-dimensional case, where $d=3$, in the
supersymmetric extension we must have two additional bosonic fields to
partner $\phi$ and $\lambda$ to form two chiral multiplets. The
form~\eqref{eq:VfromW}, can be derived from $N=1$ supergravity coupled
to two chiral superfields truncated to the $\phi$ and $\lambda$
sector. In other dimensions, the form naturally
generalises that obtained in, for example,~\cite{townsend} for the
case a single scalar field.

\subsection{Stationary points and stability}
\label{sec:solutions}

For solutions with constant $ds_{7-d}^2$ curvature, and in particular
$AdS$ space, the scalars $\phi$ and $\lambda$ are constant. Thus we need
to find the stationary points of the potential $V$. Let us introduce
the variables 
\begin{equation}
\begin{aligned}
   x &= e^{10\lambda} \\
   y &= l e^{2(d-5)\phi+2(q-p)\lambda}
\end{aligned}
\label{eq:xydef}
\end{equation}
The conditions for a stationary point of $V$ imply, in all
cases, that
\begin{equation}
   x = \frac{2y+p}{(4k/d)y^2+(p-3)}
\label{eq:xsol}
\end{equation}
together with, when $d=p=4$ (so $c\neq 0$),
\begin{equation}
   \big(y + 1\big)\left(cy^2+1\right)
         \left(cy^3 + 3cy^2 - 2y - 3\right) = 0 
\label{eq:cysol}
\end{equation}
while otherwise, i.e. when $d\neq 4$ or $p\neq 4$ (so $c=0$), 
\begin{equation}
\begin{split}
      \big(y + 1\big)\left(k(dp-2p-4d+4)y^3 
          - k(dp-10p+24)y^2 - d(p-3)(2y + 3)\right) = 0 
\end{split}
\label{eq:ysol}
\end{equation}
Note that in every case (except when $d=2p$) there is a solution
\begin{equation}
\begin{aligned}
   y &= -1 \\
   x &= \begin{cases}\
           \frac{2p}{2p-d} & \text{if $d\neq 4$ or $p\neq 4$} \\
           \frac{2}{c+1} & \text{if $d=p=4$}
        \end{cases}
\end{aligned}
\label{eq:susyxy}
\end{equation}
As we will see, these solutions correspond to the supersymmetric fixed
points as found in~\cite{gkw,gk}. Note that, since by definition,
$y/l>0$ and $x>0$, these cases always have $l=-1$ so the cycle
has negative curvature. 
%Note also that there is no supersymmetric
%solution for that case where $d=4$ and $p=2$. 
In general, since $c>0$, the remaining non-supersymmetric solutions
come from roots of the cubic factors in~\eqref{eq:cysol}
and~\eqref{eq:ysol}. 
In turns out that in all cases, the value of the potential at the
stationary point $V_0$ is negative. Thus all our solutions correspond
to $AdS$ fixed points. In general, the radius $R$ of $AdS_n$ (here
$n=7-d$) is given by 
\begin{equation}
   R^2 = -\frac{(n-1)(n-2)}{V_0}
\label{eq:R2def}
\end{equation}

One important question is whether these $AdS$ solutions are stable to
fluctuations. In the following, we do not make a full stability
analysis for fluctuations in all possible modes in the
seven-dimensional supergravity theory (\ref{Ssugra}). Instead, we
concentrate on the scalar modes $\lambda$ and $\phi$ which already
have non-trivial values in the solutions. In particular, we calculate
the mass eigenvalues of the $(\lambda,\phi)$ fluctuations about the
fixed points. It is important to recall that in $AdS$ space, instability
is characterised not by simply a negative mass-squared $M^2$, but
rather a negative mass-squared violating the Breitenlohner--Freedman
(BF) bound~\cite{bf1,bf2,mt}. For $AdS_n$ we have
\begin{equation}
   M^2 R^2 \geq -\frac{1}{4} \left(n-1\right)^2
\label{eq:BFbound}
\end{equation}
In fact, we actually considered more general fluctuations. In
particular, allowing a set of supergravity scalars
$\lambda_1,\dots,\lambda_q$ as in~\eqref{genscalaransatz}, breaking
the $SO(q)$ symmetry in the scalar field space.  We showed that this
is again a consistent truncation and that
it does in fact lead to instabilities in three cases.

%%%%%%%%%%%%%%%%%%%%%%%%%%%%%%%%%%%%%%%%%%%%%%%%%%%%%%%%%%%%%%%%%%%%%%%%%

\section{$AdS_5\times \Sigma_2$ solutions}

We start by considering the $AdS$ solutions with $d=2$, when
$\Sigma_d$ is a two-cycle. The $D=7$ gauged supergravity
metric~\eqref{metansatz} in the solutions is then given by  
\begin{equation}
   ds^2 = \frac{e^{-4\phi}R^2}{r^2}\left[
             ds^2(\bR^{1,3}) + dr^2 \right] + 
       e^{6\phi} d\bar{s}^2(\Sigma_2)
\end{equation}
where $e^{6\phi}$ and $R^2$ are constants.
Since the two-cycle has an Einstein metric it is either an
$S^2$, $\bR^2$, $H^2$ or a quotient of these spaces by a discrete group of 
isometries. Here and throughout we will not find any 
$AdS$ solutions based on flat-cycles, so we will not mention them further.
There are two cases to be considered: the
first uses the ansatz used to obtain the supersymmetric solutions
corresponding to the near horizon limits of M5-branes wrapping
K\"{a}hler two-cycles in Calabi--Yau two-folds \cite{malnun}, while
the second uses the ansatz corresponding to M5-branes wrapping K\"{a}hler
two-cycles in Calabi--Yau four-folds \cite{malnun}. Let us discuss
each case in turn. For the stability analysis, note that in this case
the BF bound is $M^2R^2\geq -4$, where $M^2$ is the mass-squared of
the fluctuation. 

\subsection{K\"{a}hler two-cycle in $CY_2$}

For this case we take $p=2,q=3$, so the scalar field ansatz
\eqref{scalaransatz} reads
\be
\Pi=\diag(e^{3\lambda},e^{3\lambda},e^{-2\lambda},e^{-2\lambda},
e^{-2\lambda})
\ee
The $SO(5)$ gauge fields are decomposed into $SO(2)\times SO(3)$ and
the $SO(3)$ gauge fields are set to zero. The $SO(2)$ gauge fields,
denoted by $B^{12}$, are determined by the $SO(2)$ spin-connection of
the two-cycle so that 
\be
B^{12}=-{1\over 4m} \bar\omega_{ab} J^{ab}
\ee
where $J$ is the K\"{a}hler structure of the two-cycle. This implies
$k=1/2$. As noted above $c=0$ for this example.

We find that the effective potential has two minima. The first occurs
when $l=-1$ and gives rise to the $AdS_5\times H^2$ supersymmetric
solution found in \cite{malnun} 
\begin{equation}
\begin{aligned}
   e^{10\lambda} &= 2 \\
   e^{6\phi} &= e^{2\lambda} \approx 1.1487 \\
   m^2 R^2 &= 2^{4/3} \approx 2.5198
\end{aligned}
\end{equation}
The second solution on the other hand is new and has 
$l=+1$. Thus it is an $AdS_5\times S^2$ solution and has 
\begin{equation}
\begin{aligned}\label{blue}
   e^{10\lambda} &\approx 6.6056 \\
   e^{6\phi} &\approx 1.1197 \\
   m^2 R^2 &\approx 1.4623
\end{aligned}
\end{equation}
This solution breaks all supersymmetry.

It is straightforward to determine the masses of the two scalar fields
about these solutions. After diagonalising the mass matrix we find that
$\phi$, $\lambda$ give rise to fluctuations with mass-squared $M^2$
satisfying: 
\begin{equation}
   M^2R^2 = -4, 12
\end{equation}
for the supersymmetric solution, and 
\begin{equation}
   M^2R^2 \approx -5.58, 22.1
\end{equation}
for the non-supersymmetric solution. Note that the BF bound is not
violated for the supersymmetric solution, as expected, but is violated
for the non-supersymmetric solution \eqref{blue}. This instability
implies that this solution cannot be the dual of a new
non-supersymmetric CFT. 

For this case, one can generalise the ansatz to include two
additional scalar fields as in  \eqref{genscalaransatz}. However, we
have checked that this leads to no further $AdS_5$ solutions. 

\subsection{K\"{a}hler two-cycle in $CY_3$}

We now have $p=4,q=1$, so the scalar field ansatz is given by
\be
\Pi=\diag(e^{\lambda},e^{\lambda},e^{\lambda},e^{\lambda},
e^{-4\lambda})
\ee
For the gauge fields we first decompose $SO(4)\to U(2)\sim U(1)\times
SU(2)$ and then let the $U(1)$ factor be determined by the spin
connection of the two-cycle. In other words we can choose a basis such
that the only non-zero components are given by 
\be
B^{12}=B^{34}=-\frac{1}{8m}\bar\omega_{ab} J^{ab}
\ee
This gives $k=1/4$ and we have $c=0$.

We again find that there are two solutions. The first is $AdS_5\times
H^2$, that is $l=-1$, with  
\begin{equation}
\begin{aligned}
   e^{10\lambda} &= \frac{4}{3} \\
   e^{6\phi} &= \frac{3}{4}e^{4\lambda} \approx 0.84147 \\
   m^2 R^2 &= \frac{9}{4} = 2.25
\end{aligned}
\end{equation}
We thus recover the supersymmetric $AdS_5\times H^2$ fixed point
first presented in \cite{malnun}. The second solution on
the other hand is again new and has $l=-1$ so the space is
$AdS_5\times H^2$, with
\begin{equation}
\begin{aligned}
   e^{10\lambda} &\approx 1.5536 \\
   e^{6\phi} &\approx 0.84824 \\
   m^2 R^2 &\approx 2.2496
\end{aligned}
\end{equation}
This solution does not preserve any supersymmetry.

The masses of the $(\phi,\lambda)$ fluctuations about these
solutions are given by
\begin{equation}
   M^2R^2 \approx -1.29, 9.29
\end{equation}
for the supersymmetric solution, and
\begin{equation}
   M^2R^2 \approx 1.40, 9.38
\end{equation}
and for the non-supersymmetric solution, so that in all cases the BF
bound is satisfied.

\section{$AdS_4\times \Sigma_3$ solutions}

These solutions are obtained when $d=3$ so $\Sigma_d$ is a three-cycle.  
The $D=7$ gauged supergravity metric for $AdS$ solutions is given by
\begin{equation}
   ds^2 = \frac{e^{-6\phi}R^2}{r^2}\left[ds^2(\bR^{1,2})+dr^2 \right] 
            +  e^{4\phi} d\bar{s}^2(\Sigma_3)
\end{equation}
where $e^{4\phi}$ and $R^2$ are constants.
Being Einstein, the three-cycle has constant curvature and for $l=1$ is
$S^3$ and for $l=-1$ is $H^3$, or a quotient of these spaces by a
discrete group of isometries. 
There are two cases to be
considered: the first uses the ansatz used to obtain the
supersymmetric solutions corresponding to M5-branes wrapping SLAG
three-cycles in Calabi--Yau three-folds~\cite{gkw} (see
also~\cite{pernicisezgin}), while the second uses the ansatz
corresponding to M5-branes wrapping associative three-cycles in
manifolds with $G_2$ holonomy~\cite{gkw}. We discuss each case in turn
and note that the BF bound now reads $M^2R^2\geq -9/4$. 

\subsection{SLAG three-cycle in $CY_3$}

In this example we have $p=3,q=2$ so the scalar field
ansatz~\eqref{scalaransatz} reads
\be
\Pi=\diag(e^{2\lambda},e^{2\lambda},e^{2\lambda},e^{-3\lambda},
e^{-3\lambda})
\ee
The $SO(5)$ gauge fields are decomposed into $SO(3)\times SO(2)$.
The $SO(2)$ gauge fields are set to zero while the $SO(3)$ gauge fields
are determined by the $SO(3)$ spin-connection of the three-cycle. Thus
if we denote the non-zero fields by $B^{ab}$ with $a,b=1,2,3$ we have 
\begin{equation}
B^{ab}=\frac{1}{2m}\bar{\omega}^{ab}
\end{equation}
This implies $k=3/8$ and again $c=0$ for this example.

We find that the effective potential has two minima, both with $l=-1$.
The first is the supersymmetric $AdS_4\times H^3$ solution of
\cite{gkw,pernicisezgin}:
\begin{equation}
\begin{aligned}
   e^{10\lambda} &= 2 \\
   e^{4\phi} &= \frac{1}{2}e^{8\lambda} \approx 0.87055 \\
   m^2 R^2 &= \sqrt{2} \approx 1.4142
\end{aligned}
\end{equation}
The second $AdS_4\times H^3$ solution is non-supersymmetric and was in
fact first found in \cite{pernicisezgin}:
\begin{equation}
\begin{aligned}
   e^{10\lambda} &= 10\\
   e^{4\phi} &\approx 1.0516 \\
   m^2 R^2 &\approx 1.3608
\end{aligned}
\end{equation}

The masses of the $(\phi,\lambda)$ fluctuations about the
supersymmetric solution are given by
\begin{equation}
   M^2R^2 \approx -1.12, 7.12
\end{equation}
while for the non-supersymmetric solution they are
\begin{equation}
   M^2R^2 \approx 0.555, 8.64
\end{equation}
These all satisfy the BF bound.

For this example we can consider adding additional scalar fields
via \p{genscalaransatz}. In particular one can take
\be
\Pi=\diag(e^{2\lambda},e^{2\lambda},e^{2\lambda},e^{-3\lambda+\alpha},
e^{-3\lambda-\alpha})
\ee
We have checked that the claim of \cite{pernicisezgin} that this
leads to an additional $AdS$ fixed point is {\it not} correct.
In particular $\alpha$ is not real.
It is interesting to note that the extra scalar field $\alpha$ 
does not violate the BF bound. In particular we find
\be
M^2R^2 = 4,68
\ee
for the supersymmetric and non-supersymmetric fixed points, respectively.

\subsection{Associative three-cycle in a manifold of $G_2$ holonomy}

For this case we have $p=4,q=1$ and the scalar field ansatz
\eqref{scalaransatz} is given by
\bea
\Pi=\diag(e^{\lambda},e^{\lambda},e^{\lambda},e^{\lambda},
e^{-4\lambda})
\eea
The non-zero $SO(5)$ gauge fields are 
taken to lie in an $SU(2)_L\subset SU(2)_L\times SU(2)_R\subset SO(5)$ 
subgroup. If we denote the non-vanishing gauge fields by $B^{mn}$, 
with $m,n=1,2,3,4$, they satisfy $B^+=0$,
and $B^-$ is determined by the $SO(3)$ spin-connection of the three-cycle:  
\begin{equation}
\begin{aligned}
  B^{-23}&=-\frac{1}{4m}\bar\omega^{23}\\
  B^{-31}&=-\frac{1}{4m}\bar\omega^{31}\\
  B^{-12}&=-\frac{1}{4m}\bar\omega^{12}
\end{aligned}
\end{equation}
This implies $k=3/16$ and again $c=0$ for this example.

We find that the effective potential has two minima, both with $l=-1$.
The first gives the supersymmetric $AdS_4 \times H_3$ solution of~\cite{agk}
with
\begin{equation}
\begin{aligned}
   e^{10\lambda} &= \frac{8}{5} \\
   e^{4\phi} &= \frac{5}{8}e^{4\lambda} \approx 0.75427 \\
   m^2 R^2 &= 2^{-11/2}5^{5/2} \approx 1.2353
\end{aligned}
\end{equation}
while the second gives a new non-supersymmetric $AdS_4 \times H_3$
solution with 
\begin{equation}
\begin{aligned}
   e^{10\lambda} &\approx 1.2839 \\
   e^{4\phi} &\approx 0.75049 \\
   m^2 R^2 &\approx 1.2362
\end{aligned}
\end{equation}

The masses of the $(\phi,\lambda)$ fluctuations satisfy
\be
M^2R^2=1.55, 6.45
\ee
for the supersymmetric solution and
\be
M^2R^2=-1.38, 6.40
\ee
for the non-supersymmetric solution and all satisfy the BF bound.

\section{$AdS_3\times \Sigma_4$ solutions}

These solutions are obtained when $d=4$ so that $\Sigma_d$ is a
four-cycle. The $D=7$ gauged supergravity metric for $AdS$ solutions is
given by 
\begin{equation}
   ds^2 = \frac{e^{-8\phi}R^2}{r^2}\left[ds^2(\bR^{1,1}) + dr^2 \right] 
              + e^{2\phi} d\bar{s}^2(\Sigma_4)
\end{equation}
where $e^{2\phi}$ and $R^2$ are constants. There are now a number of
different possibilities for the Einstein metric on $\Sigma_4$
depending on which case we are considering. In all but the case
corresponding to a M5-branes wrapping a co-associative cycle in a
$G_2$-holonomy manifold, we have $k=c$, again with a value depending
on the particular case. Note that we do not give
details of the case analogous to the M5-brane wrapping a K\"{a}hler
four-cycle in a Calabi--Yau three-fold as we do not find any $AdS_3$ 
solution, even after adding in extra scalar fields. Note that the BF
bound now reads $M^2R^2\geq -1$. 

\subsection{Co-associative four-cycle in a manifold of $G_2$ holonomy}

For this case we have $p=3$, $q=2$ so the scalar ansatz (\ref{scalaransatz})
is given by
\be
\Pi=\diag(e^{2\lambda},e^{2\lambda},e^{2\lambda},e^{-3\lambda},
e^{-3\lambda})
\ee
The non-zero gauge fields are taken to lie in $SO(3)\subset
SO(5)$. We denote them by $B^{mn}$, with $m,n=1,2,3$ and they are
determined by the anti-self-dual part of the spin connection of the
four-cycle $\omega^-$. Explicitly we let
\bea
B^{23}=-\frac{1}{m}\bar\omega^{-12}\nn
B^{31}=-\frac{1}{m}\bar\omega^{-13}\nn
B^{12}=-\frac{1}{m}\bar\omega^{-14}
\eea
This implies $k=1/3$ and we have still have $c=0$. The four-cycle is
Einstein and also conformally half-flat, so that the Weyl tensor is
self-dual. Note that for $l=1$ this means that it is either $S^4$ or
$CP^2$ if it is compact.  For $l=-1$ we denote these spaces by $C^4_-$.

We find that the effective potential has only one minima with $l=-1$.
This is the supersymmetric $AdS_3\times C^4_-$ fixed
fixed point found in \cite{gkw}:
\begin{equation}
\begin{aligned}
   e^{10\lambda} &= 3 \\
   e^{2\phi} &= \frac{1}{3}e^{8\lambda} \approx 0.80274 \\
   m^2 R^2 &= \frac{4}{9} = 0.44444
\end{aligned}
\end{equation}
The masses of the $(\phi,\lambda)$ fluctuations are given by
\begin{equation}
   M^2R^2 = 0, 40/9
\end{equation}
and so satisfy the BF bound. 

We also can consider the more general ansatz for the scalar fields
\be
\Pi=\diag(e^{2\lambda},e^{2\lambda},e^{2\lambda},e^{-3\lambda+\alpha},
e^{-3\lambda-\alpha})
\ee
However, we find no new $AdS$ fixed points. In addition we checked that
the mass of the scalar field $\alpha$ is given by
\be
M^2R^2=8
\ee
and so again these modes do not destabilise the solution. 

\subsection{SLAG four-cycle in $CY_4$} 

We now have $p=4,q=1$ so the scalar ansatz (\ref{scalaransatz}) takes
the form 
\be
\Pi=\diag(e^{\lambda},e^{\lambda},e^{\lambda},e^{\lambda},e^{-4\lambda})
\ee
The non-zero gauge fields, denoted by $B^{ab}$ with $a,b=1,2,3,4$, are
taken to lie in an $SO(4)$ subgroup of $SO(5)$ and are determined by the
$SO(4)$ spin connection: 
\be
B^{ab}=\frac{1}{2m}\bar\omega^{ab}
\ee
From~\eqref{eq:kdef} and~\eqref{eq:kc} we then have $k=c=1/3$. 
The four-cycle is Einstein and now the Weyl tensor must vanish.
In other words $\Sigma_4$ is $S^4$ for $l=1$ and 
$H^4$ for $l=-1$
or, as usual, a quotient of these spaces by a finite group of discrete
isometries.

We now find three minima of the effective potential. For $l=-1$, we
find the supersymmetric $AdS_3\times H^4$ fixed point of \cite{gkw}:
\begin{equation}
\begin{aligned}
   e^{10\lambda} &= \frac{3}{2} \\
   e^{2\phi} &= e^{-6\lambda} \approx 0.78405 \\
   m^2 R^2 &= \frac{4}{9} \approx 0.44444
\end{aligned}
\end{equation}
and also a non-supersymmetric $AdS_3\times H^4$ solution
\begin{equation}
\begin{aligned}
   e^{10\lambda} &\approx 1.2592 \\
   e^{2\phi} &\approx 0.78367 \\
   m^2 R^2 &\approx 0.44474
\end{aligned}
\end{equation}
For $l=1$ we find the non-supersymmetric $AdS_3\times S^4$ solution
\begin{equation}
\begin{aligned}
   e^{10\lambda} &\approx 3.3694 \\
   e^{2\phi} &\approx 0.23488 \\
   m^2 R^2 &\approx 0.00080517
\end{aligned}
\end{equation}

For supersymmetric solution the masses of the $(\phi,\lambda)$
fluctuations are given by
\begin{equation}
   M^2R^2 \approx 0.697, 4.30
\end{equation}
while for the non-supersymmetric solutions, with $l=-1$ we have
\begin{equation}
   M^2R^2 \approx -0.628, 4.31
\end{equation}
and for $l=1$ we get
\begin{equation}
   M^2R^2 \approx 5.24, 9.12
\end{equation}
In all cases the BF bound is satisfied.

\subsection{K\"{a}hler four-cycle in $CY_4$}

We again have $p=4,q=1$ with the scalar ansatz (\ref{scalaransatz})
of the form 
\be
\Pi=\diag(e^{\lambda},e^{\lambda},e^{\lambda},e^{\lambda},e^{-4\lambda})
\ee
The only non-vanishing $SO(5)$ gauge fields are in the $U(1)$ given by
the decomposition  $U(1)\times SU(2)\approx U(2)\subset SO(5)$. 
In this case the cycle must be Einstein and K\"{a}hler, 
denoted by $K_+$ and $K_-$
depending on whether $l=1$ or $-1$. Note that $CP^2$ and $CP^1\times CP^1$
provide examples of compact $K_+$.
The gauge fields are determined by the $U(1)$ part 
of the $SO(4)$ spin connection of the
four-cycle via the decomposition $U(1)\times SU(2)\approx U(2)\subset
SO(4)$. Explicitly, we take 
\be
B^{12}=B^{34}=-\frac{1}{8m}\bar\omega_{ab} J^{ab}
\ee
with all other gauge fields zero, where $J^{ab}$ is the K\"ahler-form of
the four-cycle. We then have $k=c=1/2$. 

We find that the effective potential has two minima. For $l=-1$ we
recover the $AdS_3\times K_-$ supersymmetric fixed point: 
\begin{equation}
\begin{aligned}
   e^{10\lambda} &= \frac{4}{3} \\
   e^{2\phi} &= e^{-6\lambda} \approx 0.84147 \\
   m^2 R^2 &= \frac{9}{16} = 0.5625
\end{aligned}
\end{equation}
and for $l=1$ we find the non-supersymmetric $AdS_3\times K_+$ solution
\begin{equation}
\begin{aligned}
   e^{10\lambda} &\approx 3.0972 \\
   e^{2\phi} &\approx 0.30836 \\
   m^2 R^2 &\approx 0.0027879
\end{aligned}
\end{equation}

For the supersymmetric solution the masses of the
$(\phi,\lambda)$ fluctuations are given by
\begin{equation}
   M^2 R^2 = 0, 40/9
\end{equation}
and for the non-supersymmetric solutions they are given by
\begin{equation}
   M^2R^2 \approx 5.16, 8.72
\end{equation}
All satisfy the BF bound. 

\subsection{Cayley four-cycle in a manifold of $Spin(7)$ holonomy}

Again we have $p=4,q=1$ with the scalar ansatz (\ref{scalaransatz})
\be
\Pi=\diag(e^{\lambda},e^{\lambda},e^{\lambda},e^{\lambda},e^{-4\lambda})
\ee
Now, however, we retain only the anti-selfdual part of the
$SO(4)\subset SO(5)$, so that the non-zero gauged fields, denoted by
$B^{ab}$ with $a,b=1,..,4$, satisfy $B^+=0$ together with  
\be
B^{-ab}=\frac{1}{2m}\bar\omega^{-ab} 
\ee
Thus, they are determined by the anti-self dual part of the $SO(4)$
spin connection, $\omega^-$, of the four-cycle. This gives $k=c=1/6$.  
In addition to being Einstein, the four-cycle must now
be conformally half-flat, which means that the Weyl tensor
is self-dual. For $l=1$ this means that it is either $S^4$ or
$CP^2$ if it is compact.  For $l=-1$ we denote these spaces by $C^4_-$.

We find three minima of the effective potential.
For $l=-1$ we find the supersymmetric $AdS_3\times C^4_-$ solution
found in \cite{gkw}
\begin{equation}
\begin{aligned}
   e^{10\lambda} &= \frac{12}{7} \\
   e^{2\phi} &= e^{-6\lambda} \approx 0.72369 \\
   m^2 R^2 &= \frac{49}{144} = 0.34028
\end{aligned}
\end{equation}
as well as the non-supersymmetric $AdS_3\times C^4_-$ solution
\begin{equation}
\begin{aligned}
   e^{10\lambda} &\approx 1.1547 \\
   e^{2\phi} &\approx 0.72346 \\
   m^2 R^2 &\approx 0.34328
\end{aligned}
\end{equation}
While with $l=1$ we find the non-supersymmetric $AdS_3\times C^4_+$ 
solution where $C^4_+$ is either $S^4$ or $CP^2$, 
\begin{equation}
\begin{aligned}
   e^{10\lambda} &= 4 \\
   e^{2\phi} &= \frac{1}{3}e^{-6\lambda} \approx 0.14509 \\
   m^2 R^2 &= 2^{-4}3^{-6} \approx 8.5734 \times 10^{-5}
\end{aligned}
\end{equation}

The masses of the $(\phi,\lambda)$ fluctuations about the 
supersymmetric solution are given by
\begin{equation}
   M^2R^2 \approx 1.89, 4.15
\end{equation}
while for the non-supersymmetric solutions, with $l=-1$ we find
\begin{equation}
   M^2R^2 \approx -1.50, 4.15
\end{equation}
and for $l=1$ we find 
\begin{equation}
   M^2R^2 \approx 5.33, 9.78
\end{equation}
All satisfy the BF bound.

\subsection{Complex-Lagrangian four-cycle in $HK_8$}

Once more we have $p=4,q=1$ and the scalar ansatz (\ref{scalaransatz})
takes the form
\be
\Pi=\diag(e^{\lambda},e^{\lambda},e^{\lambda},e^{\lambda},e^{-4\lambda})
\ee
The non-zero $SO(5)$ gauge fields lie in an $U(2)$ subgroup and are
given by the $U(2)$ spin connection of the four-cycle which must be
K\"{a}hler. Denoting the non-zero fields by $B^{ab}$ with
$a,b=1,2,3,4$ we have 
\be
B^{ab}=\frac{1}{2m}\bar{\omega}^{ab}
\ee
This gives $k=c=1/6$. In fact the Einstein K\"{a}hler cycle must
actually have constant holomorphic sectional curvature. This means
for $l=1$ we have $CP^2$ and for $l=-1$ we have the Bergmann metric $B$
on a bounded domain in $\bC^2$. As usual we can take 
quotients of these spaces and in
particular for $l=-1$ we can have compact spaces.  

The effective potential has three minima. For $l=-1$ we find
the supersymmetric $AdS_3\times B$ solution found in \cite{gk}
\begin{equation}
\begin{aligned}
   e^{10\lambda} &= \frac{6}{5} \\
   e^{2\phi} &= e^{-6\lambda} \approx 0.89638 \\
   m^2 R^2 &= \frac{25}{36} \approx 0.69444
\end{aligned}
\end{equation}
and a non-supersymmetric $AdS_3\times B$ solution:
\begin{equation}
\begin{aligned}
   e^{10\lambda} &\approx 1.3890 \\
   e^{2\phi} &\approx 0.89582 \\
   m^2 R^2 &\approx 0.69422
\end{aligned}
\end{equation}
For $l=1$ we find a non-supersymmetric $AdS_3\times CP^2$ solution:
\begin{equation}
\begin{aligned}
   e^{10\lambda} &\approx 2.9378 \\
   e^{2\phi} &\approx 0.37238 \\
   m^2 R^2 &\approx 0.0065244 
\end{aligned}
\end{equation}

The masses of the $(\phi,\lambda)$ fluctuations for
the supersymmetric solution are given by
\begin{equation}
   M^2R^2 \approx -0.419, 4.58
\end{equation}
and for the non-supersymmetric solutions, for $l=-1$
\begin{equation}
   M^2R^2 \approx 0.462, 4.55
\end{equation}
and for $l=1$ 
\begin{equation}
   M^2R^2 \approx 5.10, 8.44
\end{equation}
All satisfy the BF bound.

\subsection{SLAG four-cycle in $CY_2\times CY_2$}
\label{sec:slag4prod}

Again we take $p=4,q=1$ and the scalars are given by
\be
\Pi=\diag(e^{\lambda},e^{\lambda},e^{\lambda},e^{\lambda},e^{-4\lambda})
\ee
We first assume that the four-cycle is the product of two Einstein two-metrics 
each satisfying \p{ein}, since in this case the four-cycle is also
Einstein. Being Einstein, each two-cycle has constant curvature and
hence we are considering four-cycles of the form
$H^2\times H^2$ for $l=-1$ and $S^2\times S^2$
for $l=+1$, or a quotient thereof. 
The twisting is obtained by first decomposing $SO(5)\to SO(4)\to
SO(2)\times SO(2)$ (with ${\bf 4}\to {\bf (2,1)}+{\bf (1,2)}$ in the
last step) and identifying the $SO(2)$ factors with the spin
connections on each two cycle. Thus the only non-zero gauge fields are
given by 
\be
B^{12}=\frac{1}{2m}\bar\omega^{12} \qquad 
B^{34}=\frac{1}{2m}\bar\omega^{34}
\ee
This then gives $k=c=1$. 

The effective potential has three minima. For $l=-1$ we find
the supersymmetric $AdS_3\times H^2\times H^2$ solution found in \cite{gk}
\begin{equation}
\begin{aligned}\label{one}
   e^{10\lambda} &= 1 \\
   e^{2\phi} &= 1 \\
   m^2 R^2 &= 1
\end{aligned}
\end{equation}
as well as a non-supersymmetric $AdS_3\times H^2\times H^2$ solution:
\begin{equation}
\begin{aligned}\label{two}
   e^{10\lambda} &\approx 1.4678 \\
   e^{2\phi} &\approx 0.99497 \\
   m^2 R^2 &\approx 0.99531
\end{aligned}
\end{equation}
For $l=1$ we find a non-supersymmetric $AdS_3\times S^2\times S^2$ solution:
\begin{equation}
\begin{aligned}\label{three}
   e^{10\lambda} &\approx 2.7523 \\
   e^{2\phi} &\approx 0.48274 \\
   m^2 R^2 &\approx 0.020723
\end{aligned}
\end{equation}

The masses of the $(\phi,\lambda)$ fluctuations are now given
by 
\begin{equation}
   M^2R^2 \approx -0.828, 4.83
\end{equation}
for supersymmetric solution, while for the
non-supersymmetric solutions, for $l=-1$
\begin{equation}
   M^2R^2 \approx 1.09, 4.70
\end{equation}
and for $l=1$ 
\begin{equation}
   M^2R^2 \approx 5.02, 8.07
\end{equation}
Again all satisfy the BF bound.

For this case we can consider a more general ansatz than we have considered so
far. For the metric we take
\be
ds^2=\frac{e^{-8\phi}R^2}{r^2}\left[ds^2(\bR^{1,1})+dr^2\right] 
     +e^{2\phi}\left[e^{2h}d\bar s^2({\Sigma_2}^1) 
         + e^{-2h}d\bar s^2({\Sigma_2}^2)\right]
\ee
where have introduced a new function $e^{2h}$ which again only depends on
the coordinates on the three-space. We also allow the signs of the 
curvature of the two two-cycles, $l_1$ and $l_2$ to be in general unequal.
The scalar field ansatz is also generalised to include an extra scalar field
\be\label{twotwoscalar}
{\Pi}=\diag(e^{\lambda +\alpha},e^{\lambda+\alpha},e^{\lambda -\alpha},
         e^{\lambda-\alpha},e^{-4\lambda})
\ee
The gauge fields and the three-forms are determined as before.
We have checked that this again leads to a consistently truncated 
three-dimensional theory of gravity coupled to four scalar fields.
Let us record the effective three-dimensional Lagrangian:
\begin{equation}
   \mathcal{L} = \sqrt{-g} \Big[ R 
        - 20\left(\partial\phi\right)^2
        - 20\left(\partial\lambda\right)^2
        - 4\left(\partial h \right)^2
        - 4\left(\partial\alpha\right)^2
        - V(\phi,\lambda) \Big]
\end{equation}
where the effective potential is now given by
\bea
   m^{-2}\,V(\phi,\lambda) = &-& 2e^{-10\phi}\left(l_1e^{-2h}+l_2 
               e^{2h}\right) -\frac{1}{2}e^{-8\phi}
\left(8e^{-4\lambda}+8e^{6\lambda}\cosh(2\alpha)-e^{16\lambda}\right)\nn
&+&e^{4\lambda-12\phi}\cosh4(\alpha-h)
+\frac{1}{2}e^{-16\phi-8\lambda}
\eea 

In addition to the solutions found above we find one more, with
$l_1=1$, $l_2=-1$, i.e. $AdS_3\times S^2\times H^2$, with
\begin{equation}
\begin{aligned}\label{ping}
   e^{10\lambda} &\approx 2.4453 \\
   e^{2\alpha} &\approx 2.1367 \\
   e^{2\phi} &\approx 0.73162 \\
   e^{2h} &\approx 1.3430 \\
   m^2 R^2 &\approx 0.17815
\end{aligned}
\end{equation}

It is also interesting to determine the masses of the fluctuations
of the scalar fields $\alpha$ and $h$ for the solutions 
\p{one}, \p{two} and \p{three}. For these cases we find a block-diagonal
mass matrix when combined with $\phi,\lambda$. After diagonalising
the new block, for 
the supersymmetric $AdS_3\times H^2\times H^2$ solution \p{one} we find:
\be
M^2R^2=-0.828, 4.83
\ee
for the non-supersymmetric  $AdS_3\times H^2\times H^2$ solution \p{two}
we find
\be
M^2R^2=-1.18,5.45
\ee
which violates the BF bound, and 
for the 
non-supersymmetric  $AdS_3\times S^2\times S^2$ solution \p{three}
we find
\be
M^2R^2=-1.49, 8.33
\ee
which again violates the BF bound.
For the non-supersymmetric case \p{ping}, we find after diagonalising
the non-block diagonal $(\phi,\lambda,\alpha,h)$ mass matrix, the
following masses:
\be
M^2R^2=9.56,5.04,4.33,-1.89
\ee
which again violates the BF bound. Hence, of the product cycle cases,
it is only the supersymmetric one that does not violate the BF bound,
which is perhaps expected.

%%%%%%%%%%%%%%%%%%%%%%%%%%%%%%%%%%%%%%%%%%%%%%%%%%%%%%%%%%%%%%%%%%%%%%%%%%

\section{$AdS_2\times \Sigma_5$ solutions}
\label{sec:ads2}

Unlike the previous cases, the new solutions
with $AdS_2$ factors were found directly from
the $D=7$ equations of motion without deriving an
effective two-dimensional theory. There are two cases to
consider.

\subsection{SLAG five-cycle in $CY_5$}

The metric for this case is taken to be
\begin{equation}
   ds^2 = \frac{R^2}{r^2}\left[ -dt^2+dr^2\right] + 
            e^{2g} d\bar{s}^2(\Sigma_5)
\end{equation}
The scalar fields are taken to be trivial
\be
{\Pi_A}^i={\delta_A}^i
\ee
in order to keep the full $SO(5)$ symmetry, and 
the $SO(5)$ gauge fields are given by the $SO(5)$ spin-connection of
the five-cycle:
\be
B_{ab}=\frac{1}{2m}\bar\omega_{ab}
\ee
The five-cycle is taken to be not only Einstein,
but to have constant curvature. In other words $S^5$ for $l=1$ and 
$H^5$ for $l=-1$.
All five three-forms are now active and we find
\be
S_a=-{ {\sqrt 3}e^{-4g} \over 32}\, e^0\wedge e^r\wedge e^a
\ee
With this ansatz we only find two $AdS_2$ solutions, both are supersymmetric
and were found in \cite{gkw}. The first has $l=-1$, i.e., $AdS_2\times H^5$,
with
\begin{equation}
\begin{aligned}
   e^{2g} &= \frac{3}{4} \\
   m^2 R^2 &= \frac{9}{16} = 0.5625
\end{aligned}
\end{equation}
The second has $l=+1$, i.e., $AdS_2\times S^5$,
with
\begin{equation}
\begin{aligned}
   e^{2g} &= \frac{1}{4} \\
   m^2 R^2 &= \frac{1}{16} = 0.0625
\end{aligned}
\end{equation}

\subsection{SLAG five-cycle in $CY_3\times CY_2$}

For this case we consider an $AdS$ metric ansatz of the form
\be
ds^2=\frac{R^2}{r^2}\left[-dt^2+dr^2\right]
+e^{2g_1}d\bar s^2(\Sigma_3) +e^{2g_2}d\bar s^2(\Sigma_2)
\ee
The metric on the three-cycle and the two-cycle
are both Einstein with the signs of the curvature denoted by $l$ and $k$
respectively.
The ansatz for the scalar fields preserves $SO(3)\times SO(2)$ symmetry and
is given by
\be
\Pi=\diag(e^{2\lambda},e^{2\lambda},e^{2\lambda},
         e^{-3\lambda},e^{-3\lambda})
\ee
The $SO(5)$ gauge fields are split via $SO(5)\to SO(3)\times SO(2)$ 
and the $SO(3)$ piece $B^{ab}$, with $a,b=1,2,3$,
is determined by the spin connection on the
three-cycle, while the $SO(2)$ piece $B^{\alpha\beta}$, with $\alpha,
\beta=4,5$,
is determined by the spin connection on the two-cycle. Explicitly:
\be
B_{ab}=\frac{1}{2m}\bar\omega_{ab}\qquad  
B_{\alpha\beta}=\frac{1}{2m}\bar\omega_{\alpha\beta}
\ee
The $S$ equation of motion is solved by setting
\be
S_a=-\frac{kl}{4\sqrt{3}}e^{4\lambda-2g_1-2g_2}e^0 \wedge e^r \wedge e^a
\ee
with $S_\alpha=0$.

It is useful to define
\bea
4y&=&e^{-2g_1-2\lambda}\nn
4z&=&e^{-2g_2-2\lambda}\nn
x&=&e^{10\lambda}
\eea
We then find that the equations of motion for the above ansatz are solved
providing
\bea
40lyx&=&28x^2y^2-16z^2-3-12x+64y^2z^2x^2\nn
40kzx&=&-12x^2y^2+64z^2-3-12x+384y^2z^2x^2\nn
0&=&x-1+8z^2-4x^2y^2-32x^2y^2z^2
\eea
and we find the following three solutions.
The first has $k=l=-1$, i.e. $AdS_2\times H^3\times H^2$, and
is the supersymmetric solution found in \cite{gk}
\begin{equation}
\begin{aligned}
e^{10\lambda}&=2 \\
e^{2g_1}=e^{2g_2}&=e^{-2\lambda}\approx 0.87055 \\
m^2R^2&\approx 0.43528
\end{aligned}
\end{equation}
On the other hand for $k=-l=1$, i.e. $AdS_2\times H^3\times S^2$, we get
\begin{equation}
\begin{aligned}
e^{10\lambda}&=2 \\
e^{2g_1}=3e^{2g_2}&=e^{-2\lambda}\approx 0.87055 \\
m^2R^2&\approx0.14509
\end{aligned}
\end{equation}
or
\bea
e^{10\lambda}&\approx&0.22921\nn
e^{2g_1}&\approx&0.45751\nn
e^{2g_2}&\approx&0.95097\nn
m^2R^2&\approx&0.30432
\eea

\section{Summary}

We have found a large class of new solutions to $D=7$ gauged
supergravity that are products of $AdS_{7-d}$ space
with an Einstein space $\Sigma_d$. These can be uplifted
to obtain new solutions in $D=11$ supergravity.
We have summarised the solutions found here, as well as
the supersymmetric solutions found previously in  
\cite{malnun,agk,gkw,gk} in table~\ref{tab:table}.
\begin{table}[htbp]
\begin{center}
\begin{tabular}{|c|c|c|c|c|}
   \hline
   spacetime & embedding & cycle & supersymmetry & $m^2R^2$ \\ 
   \hline
   \hline
   $AdS_5$ & K\"{a}hler 2-cycle in $CY_2$ & $H^2$ & yes & 2.5198 \\ 
           &                              & $S^2$ & no  & 1.4623$^*$ \\
   \cline{2-5}
           & K\"{a}hler 2-cycle in $CY_3$ & $H^2$ & yes & 2.25 \\
           &                              & $H^2$ & no  & 2.2496 \\
   \hline
   $AdS_4$ & SLAG 3-cycle in $CY_3$ & $H^3$ & yes & 1.4142 \\ 
           &                        & $H^3$ & no  & 1.3608  \\
   \cline{2-5}
           & Associative 3-cycle    & $H^3$ & yes & 1.2353 \\
           &                        & $H^3$ & no  & 1.2362 \\
   \hline
   $AdS_3$ & Coassociative 4-cycle  & $C_-^4$ & yes & 0.44444 \\ 
   \cline{2-5}
           & SLAG 4-cycle in $CY_4$ & $H^4$   & yes & 0.44444 \\
           &                        & $H^4$   & no  & 0.44474 \\
           &                        & $S^4$   & no  & 0.00080517 \\
   \cline{2-5}
           & K\"{a}hler 4-cycle in $CY_4$ & $K_-^4$ & yes & 0.5625 \\
           &                              & $K_+^4$ & no  & 0.0027879 \\
   \cline{2-5}
           & Cayley 4-cycle & $C_-^4$ & yes & 0.34028 \\
           &                & $C_-^4$ & no  & 0.34328 \\
           &                & $CP^2, S^4$ & no  & 0.000085734 \\
   \cline{2-5}
           & CLAG 4-cycle in $HK_8$ & $B$    & yes & 0.69444 \\
           &                        & $B$    & no  & 0.69422 \\
           &                        & $CP^2$ & no  & 0.0065244 \\
   \cline{2-5}
           & SLAG 4-cycle in $CY_2\times CY_2$ & $H^2\times H^2$ 
                                     & yes & 1\\
           &       & $H^2\times H^2$ & no  & 0.99531$^*$ \\
           &       & $S^2\times S^2$ & no  & 0.020723$^*$ \\
           &       & $S^2\times H^2$ & no  & 0.17815$^*$ \\
   \hline
   $AdS_2$ & SLAG 5-cycle in $CY_5$ & $H^5$ & yes & 0.5625$^\dag$ \\
           &                        & $S^5$ & yes & 0.0625$^\dag$ \\
   \cline{2-5}
           & SLAG 5-cycle in $CY_2\times CY_3$ & $H^3\times H^2$ 
                                     & yes & 0.45328$^\dag$ \\
           &       & $S^2\times H^3$ & no  & 0.14509$^\dag$ \\
           &       & $S^2\times H^3$ & no  & 0.30432$^\dag$ \\
   \hline
\end{tabular}
   \caption{Table of solutions: $^*$ denotes a solution shown to be unstable,
   $C_-$ and $K_\pm$ are conformally half-flat and
   K\"{a}hler--Einstein metrics with the subscript denoting
   positive or negative scalar
   curvature and $B$ is the Bergmann metric. Note that we can also
   take quotients of all cycles by discrete groups of isometries
   and this preserves supersymmetry. $^\dag$ denotes radius in
   the seven-dimensional metric.} 
   \label{tab:table}
\end{center}
\end{table}

It would be very interesting to determine which of our new solutions
are stable, as this is a necessary condition for them to be dual to 
non-supersymmetric conformal field theories.
Our preliminary analysis involving a small number of perturbations in 
$D=7$ gauged supergravity only found that four of the
sixteen new solutions were 
unstable in the sense that the masses of the perturbations violate the 
Breitenlohner-Freedman bound \cite{bf1,bf2,mt}. Of course further
instabilities might be found when including further
perturbations of the gauged supergravity or more general
perturbations of $D=11$ supergravity.

Assuming that at least some of the new solutions are
stable, further insight into the putative dual conformal field theories could
be found by connecting them to other supersymmetric conformal
field theories by gravitational flows.
Recall that in \cite{malnun,agk,gkw,gk} supergravity solutions
were found that flowed from an $AdS_7$-type region to 
the $AdS_{7-d}\times \Sigma_d$ 
fixed points. The $AdS_7$ region, which in Poincare-type
co-ordinates has $\bR^{6-d}\times \Sigma_d$ slices, corresponds to the
UV limit describing the $(2,0)$ superconformal field theory on the wrapped
fivebranes on $\bR^{6-d}\times \Sigma_d$. The  $AdS_{7-d}\times
\Sigma_d$ fixed point at the end of the flow describes the IR physics
in $6-d$ dimensions, where one is considering length scales much
larger than the size of the cycle. 

It seems likely that there could also be supergravity solutions
that flow from the same kind of $AdS_7$ type region to 
the new $AdS_{7-d}\times \Sigma_d$ fixed points. By considering the
fall-off of the various fields, one would then be able to interpret the 
$AdS_{7-d}\times \Sigma_d$ fixed points as the IR physics arising from 
the $(2,0)$ superconformal field theory on wrapped fivebranes on
$\bR^{6-d}\times \Sigma_d$ with certain supersymmetry breaking operators
switched on. Unlike the supersymmetric flows in \cite{malnun,agk,gkw,gk},
which were found by analysing first-order BPS equations, these new
flows would be non-supersymmetric and would have to be found by solving
second-order equations.

It might also be possible to find gravity flows to or from the supersymmetric
$AdS\times \Sigma$ IR fixed points and the new non-supersymmetric 
$AdS\times \Sigma$ solutions, each with the same $\Sigma$.
By assuming a c-theorem, the direction of these flows are determined
from the $AdS$ radii as this is proportional to the central charge of 
the conformal field theory. For example, from table~\ref{tab:table},
one might look for a flow from the supersymmetric $AdS_5\times H^2$
fixed point to the non-supersymmetric $AdS_5\times H^2$ solution. On
the other hand one might look for a flow from the non-supersymmetric
$AdS_4\times H^3$ solution to the supersymmetric  $AdS_4\times H^3$
fixed point.  

Finally we note that it would also be straightforward to apply the
techniques used in this paper to find new non-supersymmetric
fixed points and flows in other supergravity duals. In particular,
one expects a variety of $AdS_2\times\Sigma_2$ fixed point solutions
for the M2-brane based on the supersymmetric ansatz in~\cite{gkpw},
similarly $AdS_3\times\Sigma_2$ and $AdS_2\times\Sigma_3$
for the D3-brane following~\cite{malnun,no} (we understand that this
is being investigated in \cite{naka}), and also new
non-supersymmetric solutions for the NS5-brane based on the
ansatze in~\cite{malnuntwo} and~\cite{gkmw,zaf,gr,gkmwtwo}. 
Note that solutions based
on different non-supersymmetric deformations of~\cite{malnuntwo}  have
recently be considered in~\cite{ass}.

%%%%%%%%%%%%%%%%%%%%%%%%%%%%%%%%%%%%%%%%%%%%%%%%%%%%%%%%%%%%%%%%%%%%%%%%%%%

\section*{Acknowledgements}

We would like to thank Ofer Aharony, Gary Gibbons and Paul Townsend
for helpful discussions. NK and SP would like to thank 
the Isaac Newton Institute for hospitality.
All authors are supported in part by PPARC
through SPG $\#$613. DW thanks the Royal Society for support. 
NK is supported by DFG.

%%%%%%%%%%%%%%%%%%%%%%%%%%%%%%%%%%%%%%%%%%%%%%%%%%%%%%%%%%%%%%%%%%%%%%%%%%%


\medskip


\begin{thebibliography}{99}

\bibitem{malnun}
J.~Maldacena and C.~Nunez,
``Supergravity description of field theories on curved manifolds and a
no go theorem,'' 
Int.\ J.\ Mod.\ Phys.\ A {\bf 16} (2001) 822
[arXiv:hep-th/0007018].
%%CITATION = HEP-TH 0007018;%%

\bibitem{malnuntwo}
J.~M.~Maldacena and C.~Nunez,
``Towards the large $N$ limit of pure $\mathcal{N} = 1$ super Yang Mills,''
Phys.\ Rev.\ Lett.\  {\bf 86} (2001) 588
[arXiv:hep-th/0008001].
%%CITATION = HEP-TH 0008001;%%

\bibitem{agk}
B.~S.~Acharya, J.~P.~Gauntlett and N.~Kim,
``Fivebranes wrapped on associative three-cycles,''
Phys.\ Rev.\ D {\bf 63} (2001) 106003
[arXiv:hep-th/0011190].
%%CITATION = HEP-TH 0011190;%%


\bibitem{no}
H.~Nieder and Y.~Oz,
``Supergravity and D-branes wrapping special Lagrangian cycles,''
JHEP {\bf 0103} (2001) 008
[arXiv:hep-th/0011288].
%%CITATION = HEP-TH 0011288;%%


\bibitem{gkw}
J.~P.~Gauntlett, N.~Kim and D.~Waldram,
``M-fivebranes wrapped on supersymmetric cycles,''
Phys.\ Rev.\ D {\bf 63} (2001) 126001
[arXiv:hep-th/0012195].
%%CITATION = HEP-TH 0012195;%%

\bibitem{Nunez:2001pt}
C.~Nunez, I.~Y.~Park, M.~Schvellinger and T.~A.~Tran,
``Supergravity duals of gauge theories from $F(4)$ gauged supergravity in six
dimensions,''
JHEP {\bf 0104}, 025 (2001)
[arXiv:hep-th/0103080].
%%CITATION = HEP-TH 0103080;%%


\bibitem{Gomis:2001vk}
J.~Gomis,
``D-branes, holonomy and M-theory,''
Nucl.\ Phys.\ B {\bf 606} (2001) 3
[arXiv:hep-th/0103115].
%%CITATION = HEP-TH 0103115;%%

\bibitem{Edelstein:2001pu}
J.~D.~Edelstein and C.~Nunez,
``D6 branes and M-theory geometrical transitions from gauged  supergravity,''
JHEP {\bf 0104}, 028 (2001)
[arXiv:hep-th/0103167].
%%CITATION = HEP-TH 0103167;%%


\bibitem{Schvellinger:2001ib}
M.~Schvellinger and T.~A.~Tran,
``Supergravity duals of gauge field theories from $SU(2)\times U(1)$ gauged
supergravity in five dimensions,''
JHEP {\bf 0106}, 025 (2001)
[arXiv:hep-th/0105019].
%%CITATION = HEP-TH 0105019;%%


\bibitem{Maldacena:2001pb}
J.~Maldacena and H.~Nastase,
``The supergravity dual of a theory with dynamical supersymmetry  breaking,''
JHEP {\bf 0109}, 024 (2001)
[arXiv:hep-th/0105049].
%%CITATION = HEP-TH 0105049;%%


\bibitem{gkpw}
J.~P.~Gauntlett, N.~Kim, S.~Pakis and D.~Waldram,
``Membranes wrapped on holomorphic curves,''
Phys.\ Rev.\ D {\bf 65} (2002) 026003
[arXiv:hep-th/0105250].
%%CITATION = HEP-TH 0105250;%%

\bibitem{Hernandez:2001bh}
R.~Hernandez,
``Branes wrapped on coassociative cycles,''
Phys.\ Lett.\ B {\bf 521} (2001) 371
[arXiv:hep-th/0106055].
%%CITATION = HEP-TH 0106055;%%


\bibitem{gkmw}
J.~P.~Gauntlett, N.~Kim, D.~Martelli and D.~Waldram,
``Wrapped fivebranes and $\mathcal{N} = 2$ super Yang-Mills theory,''
Phys.\ Rev.\ D {\bf 64} (2001) 106008
[arXiv:hep-th/0106117].
%%CITATION = HEP-TH 0106117;%%

\bibitem{zaf}
F.~Bigazzi, A.~L.~Cotrone and A.~Zaffaroni,
``$\mathcal{N} = 2$ gauge theories from wrapped five-branes,''
Phys.\ Lett.\ B {\bf 519} (2001) 269
[arXiv:hep-th/0106160].
%%CITATION = HEP-TH 0106160;%%


\bibitem{Gomis:2001vg}
J.~Gomis and T.~Mateos,
``D6 branes wrapping K\"{a}hler four-cycles,''
Phys.\ Lett.\ B {\bf 524} (2002) 170
[arXiv:hep-th/0108080].
%%CITATION = HEP-TH 0108080;%%


\bibitem{gk}
J.~P.~Gauntlett and N.~Kim,
``M-fivebranes wrapped on supersymmetric cycles. II,''
to appear in Phys. Rev. {\bf D}, hep-th/0109039.
%%CITATION = HEP-TH 0109039;%%

\bibitem{gr}
J.~Gomis and J.~G.~Russo,
``$D = 2+1$ $\mathcal{N} = 2$ Yang-Mills theory from wrapped branes,''
JHEP {\bf 0110} (2001) 028
[arXiv:hep-th/0109177].
%%CITATION = HEP-TH 0109177;%%

\bibitem{gkmwtwo}
J.~P.~Gauntlett, N.~Kim, D.~Martelli and D.~Waldram,
``Fivebranes wrapped on SLAG three-cycles and related geometry,''
JHEP {\bf 0111} (2001) 018
[arXiv:hep-th/0110034].
%%CITATION = HEP-TH 0110034;%%

\bibitem{jaume}
J.~Gomis,
``On SUSY breaking and chiSB from string duals,''
Nucl.\ Phys.\ B {\bf 624} (2002) 181
[arXiv:hep-th/0111060].
%%CITATION = HEP-TH 0111060;%%

\bibitem{DiVecchia}
P.~Di Vecchia, H.~Enger, E.~Imeroni and E.~Lozano-Tellechea,
``Gauge theories from wrapped and fractional branes,''
arXiv:hep-th/0112126.
%%CITATION = HEP-TH 0112126;%%

\bibitem{apreda}
R.~Apreda, F.~Bigazzi, A.~L.~Cotrone, M.~Petrini and A.~Zaffaroni,
``Some Comments on N=1 Gauge Theories from Wrapped Branes,''
arXiv:hep-th/0112236.
%%CITATION = HEP-TH 0112236;%%


\bibitem{bvs}
M.~Bershadsky, C.~Vafa and V.~Sadov,
``D-Branes and Topological Field Theories,''
Nucl.\ Phys.\  {\bf B463} (1996) 420,
[arXiv:hep-th/9511222].

\bibitem{pvann}
C.~N.~Pope and P.~van Nieuwenhuizen,
``Compactifications Of $D = 11$ Supergravity On Kahler Manifolds,''
Commun.\ Math.\ Phys.\  {\bf 122} (1989) 281.
%%CITATION = CMPHA,122,281;%%

\bibitem{bf1}
P.~Breitenlohner and D.~Z.~Freedman,
``Positive Energy In Anti-De Sitter Backgrounds And Gauged Extended
Supergravity,'' 
Phys.\ Lett.\ B {\bf 115} (1982) 197.
%%CITATION = PHLTA,B115,197;%%

\bibitem{bf2}
P.~Breitenlohner and D.~Z.~Freedman,
``Stability In Gauged Extended Supergravity,''
Annals Phys.\  {\bf 144} (1982) 249.
%%CITATION = APNYA,144,249;%%

\bibitem{mt}
L.~Mezincescu and P.~K.~Townsend,
``Stability At A Local Maximum In Higher Dimensional Anti-De~Sitter
Space And Applications To Supergravity,'' 
Annals Phys.\  {\bf 160} (1985) 406.
%%CITATION = APNYA,160,406;%%

\bibitem{berkone}
M.~Berkooz and S.~J.~Rey,
``Non-supersymmetric stable vacua of M-theory,''
JHEP {\bf 9901} (1999) 014
[Phys.\ Lett.\ B {\bf 449} (1999) 68]
[arXiv:hep-th/9807200].
%%CITATION = HEP-TH 9807200;%%

\bibitem{berktwo}
M.~Berkooz and A.~Kapustin,
``A comment on nonsupersymmetric fixed points and duality at large $N$,''
Adv.\ Theor.\ Math.\ Phys.\  {\bf 3} (1999) 479
[arXiv:hep-th/9903195].
%%CITATION = HEP-TH 9903195;%%

\bibitem{dggh}
F.~Dowker, J.~P.~Gauntlett, G.~W.~Gibbons and G.~T.~Horowitz,
``Nucleation of $P$-Branes and Fundamental Strings,''
Phys.\ Rev.\ D {\bf 53} (1996) 7115
[arXiv:hep-th/9512154].
%%CITATION = HEP-TH 9512154;%%

\bibitem{mms}
J.~Maldacena, J.~Michelson and A.~Strominger,
``Anti-de~Sitter fragmentation,''
JHEP {\bf 9902} (1999) 011
[arXiv:hep-th/9812073].
%%CITATION = HEP-TH 9812073;%%

\bibitem{vn}
M.~Pernici, K.~Pilch and P.~van Nieuwenhuizen,
``Gauged Maximally Extended Supergravity In Seven-Dimensions''
Phys.\ Lett.\  {\bf B143} (1984) 103.

\bibitem{vntwo}
H.~Nastase, D.~Vaman and P.~van Nieuwenhuizen,
``Consistent nonlinear KK reduction of 11d supergravity on
$AdS_7\times S_4$  and self-duality in odd dimensions'', 
Phys.\ Lett.\  {\bf B469} (1999) 96,
[arXiv:hep-th/9905075].

\bibitem{vnthree}
H.~Nastase, D.~Vaman and P.~van Nieuwenhuizen,
``Consistency of the $AdS_7\times S_4$ reduction 
and the origin of  self-duality in odd dimensions'',
Nucl.\ Phys.\  {\bf B581} (2000) 179,
[arXiv:hep-th/9911238].

\bibitem{pernicisezgin}
M.~Pernici and E.~Sezgin,
``Spontaneous Compactification Of Seven-Dimensional Supergravity Theories'',
Class.\ Quant.\ Grav.\  {\bf 2} (1985) 673.         

\bibitem{townsend}
K.~Skenderis and P.~K.~Townsend,
``Gravitational stability and renormalisation-group flow,''
Phys.\ Lett.\ B {\bf 468} (1999) 46
[arXiv:hep-th/9909070].
%%CITATION = HEP-TH 9909070;%%

\bibitem{ass}
O.~Aharony, E.~Schreiber and J.~Sonnenschein,
``Stable non-supersymmetric supergravity solutions from deformations
of  the Maldacena-Nunez background,'' 
arXiv:hep-th/0201224.
%%CITATION = HEP-TH 0201224;%%

\bibitem{naka}
M. Naka, to appear.

\end{thebibliography}
\end{document}